\documentclass[twocolumn,floats,prb,aps,showpacs]{revtex4}
\usepackage{graphicx}
\usepackage{amsfonts}
\usepackage{amsmath}
\usepackage{amssymb}
\usepackage{exscale}
\usepackage{eufrak}

\begin{document}

\title{"Moth-eaten effect" driven by Pauli blocking, revealed for Cooper
pairs}
\author{Walter V. Pogosov$^{1,2}$ and Monique Combescot$^{1}$}
\affiliation{(1) Institut des NanoSciences de Paris, Universite Pierre et Marie Curie,
CNRS, Campus Boucicaut, 140 rue de Lourmel, 75015 Paris}
\affiliation{(2) Institute for Theoretical and Applied Electrodynamics, Russian Academy
of Sciences, Izhorskaya 13, 125412 Moscow}
\date{\today }

\begin{abstract}
We extend the well-known Cooper's problem beyond one pair and study how this
dilute limit is connected to the many-pair BCS condensate. We find that, all
over from the dilute to the dense regime of pairs, Pauli blocking induces
the same "moth-eaten effect" as the one existing for composite boson
excitons. This effect makes the average pair binding energy decrease
linearly with pair number, bringing it, in the standard BCS configuration,
to half the single-pair value. This proves that, at odds with popular
understanding, the BCS gap is far larger than the broken pair energy. The
increase comes from Pauli blocking between broken and unbroken pairs.
Possible link between our result and the BEC-BCS crossover is also discussed.
\end{abstract}

\author{}
\maketitle
\date{\today }

The continuous change from the dilute to the dense regime of correlated
fermion pairs still is an open problem. Although this problem initially
arose in the context of the microscopic theory of superconductivity\cite%
{Eagles,Tonycross,Engel,Schrieffer}, its interest was recently renewed by
increasing activity in ultra-cold atomic gases. The so-called BEC-BCS
cross-over between the dilute Bose-Einstein condensate of molecules built
out of two fermion-like atoms and the dense surperfluid state of atom pairs,
is a current major question\cite{roland,reviews}. In the dilute regime,
similarities between two-atom molecules and excitons should allow their
description through a composite boson many-body formalism similar to the one
we developed for excitons \cite{Monique}. At large density, however,
excitons suffer a Mott transition to an electron-hole plasma \cite{monic}
while Cooper pairs evolve toward a BCS superconducting condensate. The
physics of this BEC-BCS crossover has also been shown to have some relevance
for Cooper pairs in high-$T_{c}$ cuprates\cite{Levin,Gantmakher}.

In this Letter, we present a conceptually trivial but yet unveiled
continuity between the Cooper's one-pair model\cite{Cooper} and the BCS
superconductivity\cite{BCS}. We do it by extending the Cooper's problem
beyond the single pair limit. We start with a "frozen" Fermi sea $\left\vert
F_{0}\right\rangle $ of noninteracting electrons and we increase the number
of electron pairs, one by one, within a layer above $\left\vert
F_{0}\right\rangle $ where the BCS potential acts. By using this approach,
we can reach the BCS regime \cite{Canon} continuously starting from the
single pair limit.

Although, at the present time, such a pair increase seems hard to
experimentally achieve, the present analysis can at least be seen as a
gedanken experiment to reveal a possible connection between two famous
problems in order to more deeply understand the role of the Pauli exclusion
principle in Cooper-paired states. This procedure can also be seen as a
simple but well-defined toy model to shed some complementary light on the
BEC-BCS crossover problem since, by changing the number of pairs, we do
change their overlap.

The extension of the Cooper's model beyond one pair faces a major many-body
problem: the exact handling of the Pauli exclusion principle between a given
number of composite particles made of fermion pairs. This can be the reason
for this extension not to have been performed yet. As proposed by Bardeen,
Cooper and Schrieffer (BCS)\cite{BCS}, the smartest way to circumvent this
difficulty is to turn to the grand canonical ensemble because the number of
fermion pairs is not fixed anymore. This procedure however masks the
existing continuity between the Cooper's problem and the dense BCS regime.
This probably is one of the reasons for Schrieffer's claim\cite{Schrieffer}
that the single-pair picture has little meaning in the dense BCS regime.

We here overcome this quite old many-body difficulty. To do so, we start
with the equations proposed by Richardson\cite{Rich} for the $N$-Cooper-pair
energy in the canonical ensemble and we manage to solve them analytically
for an \textit{arbitrary} number of pairs. This is done by extending the
method we used to solve Richardson's equations for just two pairs \cite%
{paper3}. At the present time, our mathematical approach is restricted to
the dilute limit on the single pair scale. This is why the dense limit is
here addressed by turning to the grand canonical ensemble and by extending
the BCS formalism to an arbitrary filling of the potential layer. This
allows us to show that the solution of Richardson's equations we have
obtained in the dilute limit, remains valid in the dense regime.

The result we find, proves that the\textit{\ average pair binding energy
linearly decreases with pair number over the whole density range}. For the
standard BCS configuration with a potential extending symmetrically on both
sides of the Fermi level $\left\vert F\right\rangle $ for noninteracting
electrons - a configuration which just corresponds to fill half the
potential layer - this gives an average pair binding energy reduced to half
the single pair value.

The present work also makes crystal clear how this happens. Since Pauli
blocking is the only way electrons paired by the BCS potential "interact"
(see Fig.1), the decrease of the average binding energy we find, results
plainly from the decrease of the number of states available for building
paired states within the potential layer. We can visualize this idea by
seeing each added new pair as a little moth eating one state, the number of
"moth-eaten" states increasing linearly with pair number. This "moth-eaten"
effect, which tends to decrease the effect for $N$ as compared to 1, is
actually quite standard in the many-body physics of excitons - which also
are two-fermion states.

\textbf{Ground state energy of $N$ pairs}

As stated above, our work implies handling Pauli blocking between a large
number of composite bosons. This is known to be difficult. However, our
knowledge on excitons tells us that many important features of the composite
boson many-body physics are already seen when going from $1$ to $2$ pairs:
the effect induced by Pauli exclusion principle is already present for two
pairs, in this way making the understanding for $N$ pairs far easier. This
is why, the $2$-Cooper pair problem seeming to us not out of reach, we
seriously looked for the ground state energy of $2$ pairs, with the idea to
extend the procedure to $3,4,...,N$ pairs.

\emph{(i) One-pair energy}. The energy of an electron pair with opposite
spins and zero total momentum, has been calculated by Cooper\cite{Cooper}.
It reads $\mathcal{E}_{1}=2\varepsilon _{F_{0}}-\epsilon _{c}$, where $%
\varepsilon _{F_{0}}$ is the Fermi level of the frozen sea $\left\vert
F_{0}\right\rangle $. In the weak coupling limit, the single pair binding
energy reduces to
\begin{equation}
\epsilon _{c}\simeq 2\Omega e^{-2/\rho _{0}V}
\end{equation}%
$\rho _{0}$\ is the density of states taken as constant over the potential
extension $\Omega $. Since the purpose of this Letter is to show as simply
as possible, the unrevealed consequence of Pauli blocking in BCS
superconductivity, we accept, without questioning it, the "reduced"
potential used by Bardeen, Cooper, and Schrieffer
\begin{equation}
\mathcal{V}_{BCS}=-V\sum_{\mathbf{k},\mathbf{k}^{\prime }}w_{\mathbf{k}%
^{\prime }}w_{\mathbf{k}}a_{\mathbf{k}^{\prime }\uparrow }^{\dagger }a_{-%
\mathbf{k}^{\prime }\downarrow }^{\dagger }a_{-\mathbf{k}\downarrow }a_{%
\mathbf{k}\uparrow }
\end{equation}%
$V$ is the weak potential amplitude ($\rho _{0}V\ll 1$) while $w_{\mathbf{k}%
}=1$ in the energy layer $\varepsilon _{F_{0}}<\varepsilon _{\mathbf{k}%
}<\varepsilon _{F_{0}}+\Omega $ above $\left\vert F_{0}\right\rangle
$. The main advantage of this reduced potential is to be exactly
solvable. This allows us to evidence the unrevealed physics induced
by Pauli blocking \ between Cooper pairs in a sharp way.

\begin{figure}
\vspace{-2cm}
\centerline{\scalebox{0.3}{\includegraphics{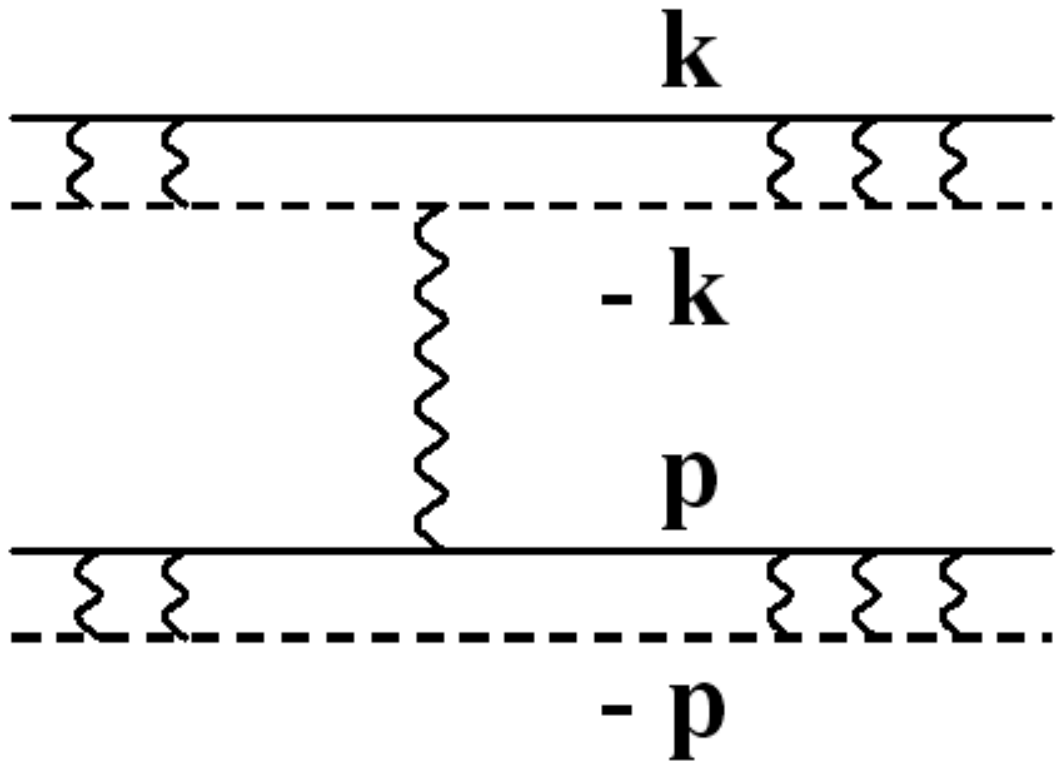}}}
\vspace{-2cm} \caption{Two Cooper pairs cannot interact through the
BCS potential given in Eq.(2): this would imply
$\mathbf{p}=\mathbf{k}$, the 2-free-pair state then reducing to zero
due to Pauli blocking.}
\end{figure}

\emph{(ii) $N$-pair eigenstates}. Forty five years ago, Richardson
has derived\cite{Rich} the \textit{exact} form for the eigenstates of
$N$ pairs. Their energies read as $\mathcal{E}_{N}=R_{1}+...+R_{N}$
where $R_{1}$, ..., $R_{N}$ are solution of $N$ algebraic equations.
For $N=2$, these equations are
\begin{equation}
1=V\sum_{\mathbf{p}}\frac{w_{\mathbf{p}}}{2\varepsilon _{\mathbf{p}}-R_{1}}+%
\frac{2V}{R_{1}-R_{2}}
\end{equation}%
%
%
%
%
%
plus a similar one with $1$ changed into $2$ - the equations for higher $N$%
's reading as Eq.(3) with all possible $R$ differences\cite{pole}.
Richardson succeeded to recover the BCS result\cite{Rich} by solving these
equations analytically in the infinite-$N$ limit for a half-filled
potential. Today, these equations are currently approached numerically for
small superconducting granules with countable number of pairs\cite{Duk}.
However, an analytical solution of these equations for arbitrary $N$ and
potential has not been given yet.

\emph{(iii) 2-pair ground state energy.} These equations actually have a
small dimensionless parameter which is the inverse of the pair number $%
N_{c}=\rho _{0}\epsilon _{c}$ above which pairs start to overlap -
this number increasing linearly with sample size. By writing these
equations in a dimensional form in terms of
$z_{i}=(R_{i}-\mathcal{E}_{1})/\epsilon _{c}$ and by performing an
expansion in $\gamma =1/N_{c}$, we found\cite{paper3}
that, for a weak coupling, the two-pair energy reads, at lowest order in $%
\gamma $ which turns out to be also an expansion in $1/\rho _{0}$
\begin{equation}
\mathcal{E}_{2}=2\left[ \left( 2\varepsilon _{F_{0}}+\frac{1}{\rho _{0}}%
\right) -\epsilon _{c}\left( 1-\frac{1}{N_{\Omega }}\right) \right]
\end{equation}%
$N_{\Omega }=\rho _{0}\Omega $ being the number of states in the potential
layer.

This result shows that Pauli blocking changes the energy of two single pairs
($2\mathcal{E}_{1}$) in two ways: It increases the free part by $1/\rho _{0}$
which just is the Fermi level change under a one-electron increase - the
extra $2$ coming from spin. It also decreases the correlated part, one state
being blocked in the 2-pair configuration. A way to better achieve this
understanding is to rewrite the single pair binding energy $\epsilon _{c}$
as
\begin{equation}
\epsilon _{c}={\rho _{0}}{\Omega }\left( \frac{2}{\rho _{0}}e^{-2/\rho
_{0}V}\right) =N_{\Omega }\epsilon _{V}
\end{equation}%
Eq. (4) then reads
\begin{equation}
\mathcal{E}_{2}=2\left[ \left( 2\varepsilon _{F_{0}}+\frac{1}{\rho _{0}}%
\right) -\left( N_{\Omega }-1\right) \epsilon _{V}\right]
\end{equation}%
Comparison between Eqs.(5) and (6) evidences that the correlation energy of
two pairs is controlled by the number of empty states $\left( N_{\Omega
}-1\right) $\ in the potential layer, i.e., the number of states available
to build the paired configuration.

\emph{(iv) $N$-pair energy in the dilute regime.} It is actually possible to
solve Richardson's equations along the same procedure as an expansion in $%
\gamma $, provided that $N/N_{c}$ stays small, a restriction which a priori
excludes the dense BCS regime, but still corresponds to $N$ arbitrary large
since only $N/N_{c}$ matters. The detailed derivation of this extension will
be presented in the long version of this Letter. Let us here give just a
sketch of our procedure.

Following Ref. \cite{paper3}, we first rewrite sums appearing in the
Richardson's equations as%
\begin{equation}
V\sum_{\mathbf{p}}\frac{w_{\mathbf{p}}}{2\varepsilon _{\mathbf{p}}-R_{i}}%
=1+\rho _{0}V\sum_{m=1}^{\infty }\frac{I_{m}}{m}z_{i}^{m}
\end{equation}%
where $I_{m}=1-e^{-2m/\rho _{0}V}$. It can then be shown that, when the
number of pairs is even, $N=2n$, the solution for the $z_{i}$'s at the
lowest order in $\gamma $ is such that
\begin{equation}
z_{1}=-z_{2n}\simeq a_{1}\sqrt{\gamma },\text{ ..., }z_{n}=-z_{n+1}\simeq
a_{n}\sqrt{\gamma }
\end{equation}%
Substitution of Eq. (8) into the Richardson's equations leads to $n$
equations for $a_{1}$, ..., $a_{n}$ which read like%
\begin{equation}
0\simeq I_{1}a_{1}+\frac{1}{a_{1}-a_{2}}+...+\frac{1}{a_{1}+a_{2}}+\frac{1}{%
2a_{1}}
\end{equation}%
We now multiply Eq. (9) by $a_{1}$ and add to similar equations for $a_{2}$,
..., $a_{n}$. This leads to%
\begin{equation}
0\simeq I_{1}\left( a_{1}^{2}+...+a_{n}^{2}\right) +n(n-1/2)
\end{equation}%
Next, we turn to the sum of Richardson's equations, as given by Eq. (7),
with \textit{two} terms kept, namely%
\begin{equation}
0\simeq I_{1}\sum_{i=1}^{n}z_{i}+\frac{I_{2}}{2}\sum_{i=1}^{n}z_{i}^{2}
\end{equation}%
Using Eqs.(10, 11), as well as the definition of $I_{1}$ and $I_{2}$, we can
find the sum of $z_{i}$ at lowest order in $\gamma $. From it, we get the
following expression for the energy of $N$-pair state
\begin{equation}
\mathcal{E}_{N}=N\left[ 2\left( \varepsilon _{F_{0}}+\frac{N-1}{2\rho _{0}}%
\right) -\epsilon _{c}\left( 1-\frac{N-1}{N_{\Omega }}\right) \right]
\end{equation}%
The same formula for $\mathcal{E}_{N}$ can be derived for an odd
number of pairs, although the form of $z_{i}$'s\ given by Eq. (8) is
somewhat more complicated.

Let us now analyze this result. The first term of $\mathcal{E}_{N}$ is equal
to twice the sum $\varepsilon _{F_{0}}$+( $\varepsilon _{F_{0}}+\frac{1}{%
\rho _{0}}$ )+ ...+($\varepsilon _{F_{0}}+\frac{N-1}{\rho _{0}}$): This just
is the energy of $N$ free free pairs added to the frozen sea $\left\vert
F_{0}\right\rangle $. The fact that we do recover the exact normal state
energy whatever $N$, can be a surprise because Eq.(12) is a priori derived
in the small $N/N_{c}$ limit. This led us to think that, most probably, the
second term of $\mathcal{E}_{N}$ also stays valid for $N$ larger than $N_{c}$%
.

\emph{(v) Energy in the dense regime.} It is first remarkable to note that
the above result exactly matches the BCS condensation energy. Indeed, this
condensation energy is known to be $E_{BCS}=\frac{1}{2}\rho _{0}\Delta ^{2}$
with $\Delta =2\omega _{c}e^{-1/\rho _{0}V}$. As $2\omega _{c}=\Omega $ is
the potential extension, $E_{BCS}$ also reads
\begin{equation}
E_{BCS}=\frac{1}{2}\rho _{0}\Omega ^{2}e^{-2/\rho _{0}V}=\frac{N_{\Omega }}{2%
}\frac{\epsilon _{c}}{2}
\end{equation}%
$N_{\Omega }/2$ is the pair number for a potential extending symmetrically
on both sides of the Fermi level. The BCS result can thus be understood as
all up and down spin electrons pairs in the potential layer form Cooper
pairs, their binding energy in this $N$-pair configuration being half the
single-pair energy: This is just Eq.(12) extrapolated to half-filling $%
N=N_{\Omega }/2$ for $N-1\simeq N$. This shows that the "moth-eaten"
effect - derived in the dilute limit - seems to stay valid in the
dense BCS regime, where pairs strongly overlap.

One important characteristic of the average binding energy we find in the
dilute limit, is its linear decrease with pair number. In order to
demonstrate the validity of this result in the dense regime, we consider
fillings different from $N_{\Omega }/2$, i.e., a potential extension
different from $\mu -\omega _{c}$ and $\mu +\omega _{c}$, the chemical
potential $\mu $ being, as usual for grand canonical ensemble, afterwards
adjusted to get the electron number. Textbook BCS formalism \cite{Tinkham}
then gives the gap equation as
\begin{equation}
1=\frac{\rho _{0}V}{2}\int_{-\mu +\varepsilon _{F_{0}}}^{\Omega -\mu
+\varepsilon _{F_{0}}}\frac{d\xi }{\sqrt{\xi ^{2}+\Delta ^{2}}}
\end{equation}%
An exact solution exists for $\mu =\varepsilon _{F_{0}}+\Omega /2$. In the
case of asymmetrical potential with boundaries still large enough to have $%
N\gg \rho _{0}\Delta $, we can replace $sinh^{-1}$ by an exponential.
Eq.(14) then gives
\begin{equation}
\Delta \simeq e^{-1/\rho _{0}V}2\sqrt{\left( \mu -\varepsilon
_{F_{0}}\right) \left( \Omega -\mu +\varepsilon _{F_{0}}\right) }
\end{equation}%
It is possible to show that the condensation energy still reads as $\frac{1}{%
2}\rho _{0}\Delta ^{2}$, with $\Delta $ now given by Eq.(15). This yields $%
N\epsilon _{c}(1-N/N_{\Omega })$, which again agrees with Eq.(12).

It can be of interest to note that, by inserting Eq.(5) into Eq.(12), we can
rewrite this condensation energy as
\begin{equation}
\mathcal{E}_{N}^{cond}=N\left( N_{\Omega }-N\right) \epsilon
_{V}=N_{occup}N_{empty}\epsilon _{V}
\end{equation}%
since $N$ is the number of occupied states in the potential layer while $%
(N_{\Omega }-N)$ is the number of empty states. This $N$ dependence makes
the condensation energy maximum when the potential acts symmetrically with
respect to the Fermi level which precisely is the BCS configuration.

A last - mathematical - result supporting the validity of Eq.(12) at large
density, is complete filling. To gain condensation energy, empty states
feeling the potential are required. There is none for complete filling. The
only possible processes then are electron exchanges. These are forbidden
within $\mathcal{V}_{BCS}$. Consequently, condensation energy must then
reduce to zero. This again agrees with Eq.(12) for $N=N_{\Omega }$.

\textbf{Physical consequences of this $N$-pair energy}

\emph{(i) Continuity between dilute and dense regimes.} The above discussion
shows that the energy of $N$ Cooper pairs given in Eq.(12), although
obtained by solving Richardson's equations in the dilute limit, remains
valid in the dense regime. This supports our understanding, reached from the
exciton many-body physics, that, due to Pauli blocking, the average pair
binding energy can only decrease when increasing the pair number, whatever
the density. It also reveals a deep connection - missed until to now -
between the Cooper's picture and the BCS regime, in spite of the fact that,
as often argued, a strong overlap between pairs in the dense regime should
destroy any link with the Cooper's model\cite{Schrieffer}. This disclosed
connection can have hidden experimental consequences in superconductivity
because, as revealed from Eq.(12), paired states do have two relevant energy
scales: the single pair energy $\epsilon _{c}$ and the excitation gap $%
\Delta $. These two quantities essentially differ by a factor of 2 in
the
exponent. This factor of 2 however is far from being unimportant because, for $%
e^{-1/\rho _{0}V}$\ very small, it makes the order of magnitude of these two
quantities quite different. Difference between the two factors has already
been noted and discussed in the literature (see, e.g., p. 169 of Ref. \cite%
{Schrieffer}).

\emph{(ii) BEC-BCS cross-over.} This connection also offers a supplementary
route to tackle BEC-BCS cross-over. Indeed, in Eagles's and Leggett's
approaches, the pair overlap is increased by decreasing the potential $V$
while we here increase this overlap by increasing $N$. These two procedures
however have some important differences: (i) By acting on $N$, the Pauli
exclusion principle blocks more and more states while this blocking stays
constant when one changes $V$ at constant potential extension $\Omega $.
(ii) Refs. \cite{Eagles,Tonycross} are based on a BCS wave function ansatz
accepted as accurate in the dense and dilute regimes but more questionable
along the crossover\cite{Tonycross}. In contrast, we here use the \textit{%
exact} wave function obtained by Richardson for the ground state energy of $N
$ pairs. In spite of these differences, the general conclusion of Ref. \cite%
{Tonycross} and the present letter stays the same: ground state pairs in the
dilute and dense regimes are not so much different, a conclusion at odds
with Schrieffer's claim \cite{Schrieffer}.

\emph{(iii) Excitation gap.} Since the average pair binding energy decreases
over the whole density range, the reader most probably stays with one major
question: what controls the gap in the excitation spectrum of
superconductors? The answer again is Pauli blocking. When a pair is broken,
the system not only looses its binding energy, but all the remaining
unbroken pairs have their average binding energy decreased: the two free
electrons resulting from the Cooper pair broken by a photon, block two pair
states (the photon momentum being small but not exactly zero). The remaining
unbroken pairs feel these blocked states when trying to construct their
correlated state. The latter effect increases with the number of unbroken
pairs to end in the dense regime, by being far larger than the broken pair
energy.

Preliminary results show strong indications that when $N$ becomes larger
than $N_{c}$, the threshold energy to break a pair achieves the same $V$ and
$N$ dependences as $\Delta $. Similar result for the gap change from
single-pair to a more cooperative regime was actually found in Refs. \cite%
{Eagles,Tonycross} within a variational BCS-like approach, this change going
along a weak singularity\cite{Engel}.

\emph{(iv) Superfluid and virtual pairs.} We here deal with paired
states formed out of all the $2N$ up and down spin electrons added in
the energy layer where the potential acts. These electrons feel the
potential; they are correlated and form the $N$ pairs we consider in
this Letter. These pairs are the ones which are "condensed" into the
same quantum-mechanical state in the BCS wave function ansatz.
Schrieffer calls them\cite{Schrieffer} "superfluid pairs".

These "superfluid pairs" have to be contrasted with what Schrieffer\cite%
{Schrieffer} calls "virtual pairs". The latters correspond to "electrons
excited above the Fermi level" $\left\vert F\right\rangle $ of the \textit{%
noninteracting} electrons. It is of importance to note that the concept of
"virtual pairs" is physically relevant in the dense regime only because the
Fermi level $\left\vert F\right\rangle $ must not be smeared out too much by
interactions in order to keep some physical meaning. As a result, the
understanding of the BCS regime in terms of "virtual pairs" tends to break
in an artificial way a possible continuity with the dilute limit.

These "virtual pairs" are the ones commonly used to give a qualitative
understanding\cite{Tinkham,LP} to the BCS condensation energy, when writing
it as a pair number multiplied by a pair energy. Indeed, their number,
deduced from the width of the BCS distribution change, is of the order of $%
N_{\Delta }=\rho _{0}\Delta $. This gives a pair energy of the order of ${%
\Delta }$, within an irrelevant factor of 2. From it, it is then concluded
\cite{Fetter} that the "pair energy" must be of the order of the gap. It is
clear that this conclusion fully relies on what is chosen as pair number. By
instead taking the total number of pairs $N_{\Omega }/2$ feeling the
potential, as we here do - this number being the natural pair number of the
problem - the \textit{same} BCS condensation energy gives a pair energy
exactly \textit{equal} to $\epsilon _{c}/2$ in agreement with Eq. (12).

We wish to stress that, when compared to the understanding based on "virtual
pairs", understanding based on"superfluid pairs" provide a natural
connection between the dilute and dense regimes of pairs. Within these
"superfluid pairs", the large value of the excitation gap is due to
many-body effects arising from Pauli blocking between broken and unbroken
pairs, these many-body effects definitely having some physical relevance.

\textbf{Conclusion}

We have extended the well-known Cooper's model beyond the one-pair
configuration and revealed the simple link which exists between this model
and BCS superconductivity. We show that the average pair binding energy
\textit{linearly decreases with pair number}. In agreement with our
understanding of the exciton many-body physics, the Pauli exclusion
principle induces a "moth-eaten effect" on Cooper pairs, unveiled here for
the first time. The average pair binding energy in the standard BCS
configuration is shown to only be half the single pair value, as a result of
their mutual Pauli blocking. This makes the excitation gap in the dense
regime far larger than the broken pair energy. This increase is due to the
Pauli exclusion principle induced by many-body effects between broken and
unbroken pairs. Our work evidences that superconductors have a hidden second
energy scale - the average pair binding energy - which, in the weak coupling
limit, is far smaller than the gap. This result should stimulate new
experiments in this very old field. Finally, to precisely understand how the
isolated pair and BCS regimes are connected, can be very valuable in a
possible approach to the BEC-BCS cross-over within a single composite boson
many-body formalism \cite{Monique}.

M.C. wishes to thank Tony Leggett for enlightening discussions during her
invitation by the University of Illinois at Urbana-Champaign. W. V. P.
acknowledges supports from the French Ministry of Education during his stay
in Paris, RFBR (project no. 09-02-00248), and Dynasty Foundation. We also
have received very many valuable advises from Nicole Bontemps, Tristan Cren
and Dimitri Roditchev.


\begin{thebibliography}{99}
\bibitem{pole} While this sum is real, the $R_i$'s are complex, so that
possible poles at $2\epsilon _{p}$ are avoided.
\end{thebibliography}
\end{document}